\documentclass{article}
\usepackage{amsmath,amsthm,amssymb,bm,lscape}

\newcommand{\sn}{{\rm sn}}
\newcommand{\cn}{{\rm cn}}
\newcommand{\dn}{{\rm dn}}
\newcommand{\s}{{\hat{S}}}
\newcommand{\bs}{{\bar{S}}}
\newcommand{\Q}{{\hat{Q}}}

\def\tSs{\tilde{{\mathsf S}}}

\def\Ths{{\mathsf \Theta}}
\def\Ss{{\mathsf S}}
\def\Hs{{\mathsf H}}
\def\Ks{{\mathsf K}}

\def\Qs{{\mathsf Q}}

\newcounter{app}
\newcommand{\app}[1]{
\refstepcounter{app}{\vspace{7mm}
\noindent\Large\bf Appendix
\theapp.
 \ #1 \par \vspace{5mm}}
\setcounter{equation}{0}
\def\theequation{\Alph{app}.\arabic{equation}}}
\begin{document}
\title{A new $Q$-matrix in the Eight-Vertex Model}
\author{Klaus Fabricius,\\Physics Department, University of Wuppertal,
42097 Wuppertal, Germany
\footnote{e-mail Fabricius@theorie.physik.uni-wuppertal.de}}
\maketitle
\begin{abstract}
\noindent
We construct a $Q$-matrix for the eight-vertex model at roots of unity for crossing parameter $\eta=2mK/L$ with
odd $L$, a case for which the existing constructions do not work. 
The new Q-matrix $\Q$  depends on the spectral parameter $v$ and also on a free parameter $t$. 
For $t=0$ $\Q$ has the standard properties.
For $t\neq 0$, however, it does not commute with the operator $S$ and not with itself for different values
of the spectral parameter. We show that the six-vertex limit of $\Q(v,t=iK'/2)$ exists.
\end{abstract}
\noindent
An essential tool in Baxter's solution of the eight-vertex model \cite{bax0,bax1,bax2,bax3}
is the $Q$-matrix which  satisfies the $TQ$ equation
\begin{equation}
T(v)Q(v)=[\rho h(v-\eta)]^NQ(v+2\eta) +[\rho h(v+\eta)]^NQ(v-2\eta)
\label{funeqn}
\end{equation}
and commutes with $T$.
Here $T(v)$ is the transfer matrix of the eight-vertex model (\ref{t8}).
Combined with periodicity properties of $Q(v)$ in the complex $v$-plane equ. (\ref{funeqn}) leads
to the derivation of Bethe's equations and the solution of the model.
For generic values of the crossing parameter $\eta$ the transfer matrix $T$ has a non 
degenerate spectrum. For rational values of $\eta/K$ however
this is not the case. This leads to the existence of different $Q$-matrices which all 
satisfy equ. (\ref{funeqn}).
In ref.\cite{bax0} Baxter constructs a $Q$-matrix valid for 
\begin{equation}
2L\eta = 2m_1K + im_2K'
\end{equation}
with integer $m_1,m_2,L$. In ref.\cite{bax1} Baxter derived a $Q$-matrix valid
for generic values of $\eta$.
As these $Q$-matrices are different we distinguish them by writing $Q_{72}$ and
$Q_{73}$ respectively for the constructions in ref. \cite{bax0} and ref. \cite{bax1}.
It turned out, however, that $Q_{72}$ 
has interesting properties beyond its role in equ. (\ref{funeqn})
{\it because} of its restriction to rational values of $\eta/K$.\\
In ref. \cite{Newdev} it is conjectured that $Q_{72}(v)$ satisfies the following
functional relation:\\
For $N$ even and $\eta=m_1K/L$ where either $L$ even or $L$ and $m_1$ odd  
\begin{eqnarray}
& &e^{-N\pi i v/2K}Q_{72}(v-iK')\nonumber\\
&=&A\sum_{l=0}^{L-1}h^N(v-(2l+1)\eta)
\frac{Q_{72}(v)}{Q_{72}(v-2l\eta)Q_{72}(v-2(l+1)\eta)}
\label{con}
\end{eqnarray}
$A$ is a normalizing  constant matrix independent of $v$ 
that commutes with $Q_{72}$ and $h(v) = H(v)\Theta(v)$.
There is a proof of this conjecture valid for $L=2$ in ref.\cite{fusion}.
This functional relation is important as it allows the conclusion that the dimension 
of eigenspaces of degenerate eigenvalues of the $T$-matrix is a power of 2, a result also
true in the six-vertex model provided the roots of the Drinfeld polynomial of the loop algebra symmetry
are \mbox{distinct \cite{Odyssey}.}\\
The reason why the case $L$ odd and $m_1$ even is excluded in (\ref{con}) is that $Q_{72}$
does not exist in this case \cite{Newdev}.\\
The purpose of this paper is to close this gap\footnote{There exists now a related investigation by Roan \cite{Roan}}
. We construct for even $N$ a $Q$-matrix
which exists for $\eta=2mK/L$ for odd $L$ which satisfies the functional
relation (\ref{con}). Beyond that we shall show that  for $\eta=2mK/L$ a more general $Q$-matrix
exists which depends on a free parameter $t$ and which does not commute with $R$ and $S$
where
\begin{equation}
 R =  \underbrace{\sigma_1 \otimes \sigma_1 \otimes \cdots \otimes \sigma_1}_{\rm N~~ factors}
\hspace {0.6 in} 
S = \underbrace{\sigma_3 \otimes \sigma_3 \otimes \cdots \otimes \sigma_3}_{\rm  N~~ factors}
\label{SRop}
\end{equation}
and not even with itself for different spectral parameters
\[
[Q(v_1,t),Q(v_2,t)] \neq 0
\]
This phenomenon has also been observed by Bazhanov and Stroganov in ref. \cite{BazStrog}
for their column-to-column transfer matrix $T_{col}$ which acts like a $Q$-matrix 
in the six-vertex model: It satisfies (\ref{funeqn}) and it commutes with $T_6$. 
But it does not commute with itself for different arguments.\\
We use the notation of Baxter's 1972 paper. We denote our new $Q$-operators by 
$\Q_R$,$\Q_L$ and $\Q$. They depend on two arguments $v$ and $t$, e.g $\Q_R(v,t)$.
For brevity we shall write $\Q_R(v)$ instead of $\Q_R(v,0)$. 
The symbols $Q$, $Q_R$ etc. refer to all types of $Q$ matrices.\\
The plan of this paper is as follows.
In section 1 we describe the various steps in the construction of $Q$. We
first outline  in section 1.1 the general method developed by Baxter and his
solution leading to $Q_{72}$. In section 1.2 we present our new $\Q_R$ operator
and describe its range of validity. In section 1.3 we introduce the matrix $Q_L$ and show 
in section 1.4 that the famous equation \mbox{$Q_L(u)Q_R(v)$ = $Q_L(v)Q_R(u)$} which Baxter proved
for $Q_{72}$ and $Q_{73}$ is also satisfied by $\Q(v)$ . In section 2 we study the quasiperiodicity
properties of $Q(v)$ and show that there exists a link between quasiperiodicity of $Q_R$ with
quasiperiod $iK'$ and non existence of $Q_R^{-1}$. We summarize in section 3 the properties of
$\Q(v)$ and describe in section 4 the exotic properties of $\Q(v,t)$ for $t \neq 0$.

\section{Construction of a $Q$-matrix for $\eta=2mK/L$.}
\subsection{Baxter's construction of $Q_{72}$.}
The goal is to find a matrix $Q_R$ of the form
\begin{equation}
[Q_R(v)]_{\alpha |\beta}={\rm Tr}S_R(\alpha_1, \beta_1)
S_R(\alpha_2, \beta_2)\cdots S_R(\alpha_N, \beta_N)
\label{TrSR}
\end{equation}
where $\alpha_j$ and $\beta_j=\pm 1$ and $S_R(\alpha,\beta)$ is a
matrix of size $L\times L$ such that $Q_R$ satisfies 
\begin{equation}
T(v)Q_R(v)=[\rho h(v-\eta)]^N Q_R(v+2\eta) +[\rho h(v+\eta)]^NQ_R(v-2\eta)
\label{TQReqn}
\end{equation}
The $Q$-matrix occurring in equ. (\ref{funeqn}) is then
\begin{equation}
Q(v) = Q_R(v)Q^{-1}_R(v_0)
\label{QRQRinv}
\end{equation}
for some constant $v_0$. Therefore it is necessary that $Q_R(v)$ is a regular matrix.
The problem to construct a $Q_R$ of the form (\ref{TrSR}) satisfying (\ref{TQReqn}) 
is posed and solved by Baxter in Appendix C of ref. \cite{bax0}.
In order to construct a $Q_R$-matrix which is regular for $\eta=mK/L$ for even $m$ and odd $L$ 
we shall search for other solutions of Baxter's fundamental equations.\\
These equations are (see (C10), (C11) in ref. \cite{bax0})\\
\begin{eqnarray}
&&(ap_n-bp_m)S_R(+,\beta)_{m,n} + (d-cp_mp_n)S_R(-,\beta)_{m,n} =0\nonumber \\
&&(c-dp_mp_n)S_R(+,\beta)_{m,n} + (bp_n-ap_m)S_R(-,\beta)_{m,n} =0\nonumber \\
\label{sys}
\end{eqnarray}
where $\beta=+,-$,  $m,n=1,\cdots L$ and $a,b,c,d$ are defined in (\ref{bw}).
Equs. (\ref{sys}) determine the elements of the local matrices $S_R(\alpha,\beta)$ occurring in (\ref{TrSR})
provided that the determinant of this system of homogeneous linear equations vanishes:
 \begin{equation}
(a^2+b^2-c^2-d^2)p_mp_n=ab(p_m^2+pn^2_n)-cd(1+p_m^2p_n^2)
\label{det}
\end{equation}
This determines $p_n$ if $p_m$ is given. Setting 
\begin{equation}
p_m=k^{1/2}\sn(u)
\label{pm}
\end{equation}
it follows from (\ref{snumv})
that
\begin{equation}
p_n=k^{1/2}\sn(u\pm 2\eta)
\label{pn}
\end{equation}
Baxter selected a solution which has  non vanishing diagonal elements
$S_R(\alpha,\beta)_{0,0}$ and $S_R(\alpha,\beta)_{L,L}$.
In order to allow $S_R(\alpha,\beta)_{m,n}$ to have non vanishing diagonal elements
$S_R(\alpha,\beta)_{0,0}$ and $S_R(\alpha,\beta)_{L,L}$
equ. (\ref{det}) has to be satisfied for $n=m$. Then
\begin{equation}
\sn(u) = \sn(u \pm 2\eta)
\end{equation}
This fixes the parameter $u$ to become $u=K \pm \eta$ and leads to the restriction to discrete $\eta$:
\begin{equation}
2L\eta=2m_1K+im_2K'
\label{rou0}
\end{equation}
One obtains from (\ref{pm}) and (\ref{pn}) that
\begin{equation}
p_n=k^{1/2}\sn(K+(2n-1)\eta)
\label{pnn}
\end{equation}
and from (\ref{sys})
\begin{eqnarray}
&&  {\Ss_R(\alpha,\beta)(v)_{k,l}} = \nonumber \\
&&\delta_{k+1,l} {  u^{\alpha}(v+K-2k\eta)\tau_{-k,\beta}}
                                    +\delta_{k,l+1} {  u^{\alpha}(v+K+2l\eta)\tau_{l,\beta}}+ \nonumber \\
&&\delta_{k,1}\delta_{l,1}{  u^{\alpha}(v+K)\tau_{0,\beta}} +
                                    \delta_{k,L}\delta_{l,L}{  u^{\alpha}(v+K+2L\eta)\tau_{L,\beta}} \nonumber \\
\label{SR}
\end{eqnarray}

for $1<k\leq L$, $1<l\leq L$ and
where
\begin{equation}
u^{+}(v) = \Hs(v) \hspace{0.5 in} u^{-}(v) = \Ths(v)
\label{ualpha}
\end{equation}
if
\begin{equation}
\eta=m_1 K/L.
\end{equation}
$Q_{R,72}$ is the matrix $Q_R$ defined in (\ref{TrSR}) with $S_R$ given by (\ref{SR}).\\
It has been shown in \cite{Newdev} that $Q_R$ based on (\ref{SR}) is singular if $m_1$ is even and $L$ is odd.
In the following subsection we show that an alternative construction leads for these $\eta$-values to a 
regular $Q_R$-matrix.
\subsection{Another $Q$-matrix}
\noindent
To obtain another solution $\s_R$ of (\ref{sys}) and (\ref{det}) we consider the possibility 
that the elements of $\s_R(\alpha,\beta)_{m,n}$
form cycles
\[
\s_R(\alpha,\beta)_{1,2},\s_R(\alpha,\beta)_{2,3},\cdots,\s_R(\alpha,\beta)_{L-1,L},\s_R(\alpha,\beta)_{L,1}
\]
and 
\[
\s_R(\alpha,\beta)_{2,1},\s_R(\alpha,\beta)_{3,2},\cdots,\s_R(\alpha,\beta)_{L,L-1},\s_R(\alpha,\beta)_{1,L}
\]
instead of imposing the condition that  $\s_R(\alpha,\beta)_{m,n}$ has two diagonal elements.
In this case a set of functions $p_n$ consistent with (\ref{pm}) and (\ref{pn}) is
\begin{equation}
p_n = k^{1/2}\sn(t+(2n-1)\eta)
\label{newp}
\end{equation}
From the condition that
\begin{equation}
\s_R(\alpha,\beta)_{L,L+1}=\s_R(\alpha,\beta)_{L,1}
\end{equation}
if follows that $p_1$ and $p_L$ must have arguments which differ by $2\eta$:
\begin{equation}
\sn(t+(2L-1)\eta) = \sn(t+\eta-2\eta)
\end{equation}
This is satisfied if
\begin{equation}
2L \eta = 4mK+2im_2K'
\label{rou2}
\end{equation}
This condition differs from (\ref{rou0}).
The solution of equs. (\ref{sys}) with the set of $p_n$-functions  (\ref{newp}) as input is
\begin{equation} 
  { \s_R(\alpha,\beta)_{k,l} =  \delta_{k+1,l}w^{\alpha}(v-t-2k\eta)\tau_{\beta,-k}
                            +\delta_{k,l+1}u^{\alpha}(v+t+2l\eta)\tau_{\beta,l}}
\label{SRkl}
\end{equation}
with $u^{\alpha}$ defined in equ. (\ref{ualpha}) and $w^{\alpha}$ is given by
\begin{equation}
w^{+}(v) = -\Hs(v) \hspace{0.5 in} w^{-}(v) = \Ths(v)
\label{walpha}
\end{equation}
Note that the first component of $w^{\alpha}$ differs from $u^{+}$ by a minus sign.\\
We consider only the case $m_2=0$ in (\ref{rou2}).
Then 
\begin{equation}
\eta = 2mK/L
\label{rou3}
\end{equation} 
We shall denote the $Q_R,Q_L$ and $Q$-matrices derived from $\s_R,\s_L$ by $\Q_R,\Q_L$ and $\Q$. 
We distinguish the following cases:\\
1.\\
If $L$ is odd the resulting $\Q$-matrices cover exactly the set of discrete $\eta$-values which is missing in the 
original solution (\ref{SR})-(\ref{ualpha}).
We note that for $t=K$ this solution becomes identical to case (\ref{SR})-(\ref{ualpha}) with singular $\Q_R$.
But for generic $t$ (especially $t=0$) { $ \Q_R$ is regular}. It must be stressed, however, that the
regularity has not been proved analytically but numerically for sufficiently large systems to allow the
occurrence of degenerate eigenvalues of the transfer matrix $T$. See also appendix C of ref. \cite{bax0}
and ref. \cite{Newdev}.\\
2.\\
$L$ is even but both $L_1=L/2$ and $m$ are odd. \\
Then $\eta=mK/L_1$ is that set of $\eta$-values for which
the solution (\ref{SR})-(\ref{ualpha}) leads to regular $Q_R$ matrices. It turns out that in this
case the $\Q_R$-matrix resulting from solution (\ref{SRkl}) is singular.\\
3.\\
$L$ and $L/2$ are even and $m$ is odd.\\
In this case both solutions (\ref{SR})-(\ref{ualpha}) and (\ref{SRkl}) give regular
$Q_R$-matrices. But the matrices $\s_R(\alpha,\beta)$ differ in size by a factor 2.

The conclusion is that the two sets of $Q$-matrices (\ref{SR})-(\ref{ualpha}) and (\ref{SRkl}) 
are complementary in the sense that for $\eta=mK/L$ and odd $L$ what is missing in the first set is present in the second and vice versa.
\subsection{The matrix $\Q_L$.}
\noindent
To get finally a $Q$-matrix which commutes with the transfer matrix $T$ and satisfies equ. (\ref{funeqn}) 
Baxter introduced a second matrix $Q_L$.
By transposing equ. (\ref{TQReqn}) and replacing $v$ by $-v$ one obtains
\begin{equation}
Q_L(v)T(v)=[\rho h(v-\eta)]^N Q_L(v+2\eta) +[\rho h(v+\eta)]^NQ_L(v-2\eta)
\label{funeqnL}
\end{equation}
with
\begin{equation}
Q_L(v) = Q_R^t(-v)
\label{QL}
\end{equation}
and
\begin{equation}
[Q_L(v)]_{\alpha |\beta}={\rm Tr}S_{L}(\alpha_1, \beta_1)
S_{L}(\alpha_2, \beta_2)\cdots S_{L}(\alpha_N, \beta_N)
\label{TrSL}
\end{equation}
We perform this construction for the new $\Q_R$ matrix.
The local matrices $\s_L$ are obtained from (\ref{SRkl})
\begin{equation}
  { \s_L(\alpha,\beta)_{k,l}(v) =  \s_R(\beta,\alpha)_{k,l}(-v)}
\label{SLkldef}
\end{equation}
\begin{equation} 
  { \s_L(\alpha,\beta)_{k,l} =  \delta_{k+1,l}\tau_{\alpha,-k}u^{\beta}(v+t+2k\eta)
                            +\delta_{k,l+1}\tau_{\alpha,l}w^{\beta}(v-t-2l\eta)}
\label{SLkl}
\end{equation}
\subsection{The relation $Q_L(u)Q_R(v)=Q_L(v)Q_R(u)$.}
\noindent
To prove that the $Q$-matrix defined by
\begin{equation}
Q(v) = Q_R(v)Q_{R}^{-1}(v_0)
\label{Q}
\end{equation}
commutes with the transfer matrix $T$ Baxter shows in ref. \cite{bax0} that the relation
\begin{equation}
Q_L(v)Q_R(u)=Q_L(u)Q_R(v)
\label{QLQR}
\end{equation}
holds. 
Then
\begin{equation}
Q(v) = Q^{-1}_L(u)Q_L(v) =  Q_R(v)Q^{-1}_R(u) 
\end{equation}
commutes with $T(v)$.
To prove (\ref{QLQR}) it is shown in ref.\cite{bax0} that
$S_{L}(\alpha,\gamma)_{m,n}(u)S_{R}(\gamma,\beta)_{m'n'}(v)$ and $S_{L}(\alpha,\gamma)_{m,n}(v)S_{R}(\gamma,\beta)_{m'n'}(u)$ 
are related by a similarity transformation.
\begin{equation}
S_{L}(\alpha,\gamma)_{m,n}(u)S_{R}(\gamma,\beta)_{m'n'}(v)=
Y_{m,m';k,k'}S_{L}(\alpha,\gamma)_{k,l}(v)S_{R}(\gamma,\beta)_{k'l'}(u)Y^{-1}_{l,l';n,n'} 
\label{YS}
\end{equation}
with diagonal matrix $Y$
\begin{equation}
Y_{m,m';k,k'}=y_{m,m'}\delta_{m,k}\delta_{m',k'}
\label{Ydiag}
\end{equation}
To investigate whether the matrices $\Q_R$ and $\Q_L$ defined in (\ref{TrSR}),(\ref{SRkl}) and (\ref{TrSL}),(\ref{SLkl})
fulfill such a relation we define a series of abbreviations. According to (\ref{SRkl}) we write
\begin{equation}
\s_{R}(\alpha,\beta)_{m,n}=\Phi^{\alpha}_{m,n}\bar{\tau}^{\beta}_{m,n}
\label{S_R}
\end{equation}
where 
\begin{equation}
\Phi^{\alpha}_{m,n} =  \epsilon^{\alpha}_{m,n}f^{\alpha}(v_{m,n})
\label{Phi}
\end{equation}
\begin{equation}
v_{m,n} = \delta_{m-1,n}(v+t+2n\eta) +\delta_{m+1,n}(v-t-2m\eta)
\label{xmn}
\end{equation}
\begin{equation}
\epsilon^{\alpha}_{m,n}=  \delta_{m-1,n} -\alpha \delta_{m+1,n} \hspace{0.6 in} \alpha = \pm 1
\label{eps}
\end{equation}
\begin{equation}
\bar{\tau}^{\beta}_{m,n}=  \delta_{m-1,n}\tau_{\beta,n}+ \delta_{m+1,n}\tau_{\beta,-m} 
\label{taub}
\end{equation}

and $f^{+}(v) = H(v)$, $f^{-}(v) = \Theta(v)$,
 $\delta_{m+L,n} = \delta_{m,n}$.\\
Equivalently we write following (\ref{SLkl})
\begin{equation}
\s_{L}(\alpha,\beta)_{m,n}=\tau^{'\alpha}_{m,n}\chi^{\beta}_{m,n}
\label{S_L}
\end{equation}
where 
\begin{equation}
\chi^{\beta}_{m,n} =  \lambda^{\beta}f^{\beta}(u_{m,n})
\label{chi}
\end{equation}
\begin{equation}
u_{m,n} = \delta_{m-1,n}(v-t-2n\eta) +\delta_{m+1,n}(v+t+2m\eta)
\label{ymn}
\end{equation}
\begin{equation}
\lambda^{\beta}_{m,n}= -\beta \delta_{m-1,n} + \delta_{m+1,n}
\label{lambda}
\end{equation}
\begin{equation}
\bar{\tau'}^{\alpha}_{m,n}=  \delta_{m-1,n}\tau'_{\alpha,n}+ \delta_{m+1,n}\tau'_{\alpha,-m} 
\label{taubp}
\end{equation}
It follows then from (\ref{S_R}) and (\ref{S_L})
\begin{equation}
\s_{L}(\alpha,\gamma)_{m,n}(u)\s_{R}(\gamma,\beta)_{m'n'}(v) = 
\tau^{'\alpha}_{m,n}\chi^{\gamma}_{m,n}(u)\Phi^{\gamma}_{m',n'}(v)\tau^\beta_{m',n'}
\end{equation}
																						      and from (\ref{Phi}) and (\ref{chi}) one obtains
\begin{eqnarray}
&&\chi^{\gamma}_{m,n}(u)\Phi^{\gamma}_{m',n'}(v)= \nonumber \\
&&(\delta_{m+1,n}\delta_{m'+1,n'}+\delta_{m-1,n}\delta_{m'-1,n'})(\Theta(u_{m,n})\Theta(v_{m',n'})-H(u_{m,n})H(v_{m',n'}))+\nonumber \\
&&(\delta_{m+1,n}\delta_{m'-1,n'}+\delta_{m-1,n}\delta_{m'+1,n'})(\Theta(u_{m,n})\Theta(v_{m',n'})+H(u_{m,n})H(v_{m',n'}))
\end{eqnarray}
with non vanishing elements
\begin{equation}
\chi^{\gamma}_{m,m+1}(u)\Phi^{\gamma}_{m',m'+1}(v)=\Theta(u_{m,m+1})\Theta(v_{m',m'+1})-H(u_{m,m+1})H(v_{m',m'+1})
\label{chipp}
\end{equation}
\begin{equation}
\chi^{\gamma}_{m,m-1}(u)\Phi^{\gamma}_{m',m'-1}(v)=\Theta(u_{m,m-1})\Theta(v_{m',m'-1})-H(u_{m,m-1})H(v_{m',m'-1})
\label{chimm}
\end{equation}

\begin{equation}
\chi^{\gamma}_{m,m+1}(u)\Phi^{\gamma}_{m',m'-1}(v)=\Theta(u_{m,m+1})\Theta(v_{m',m'-1})+H(u_{m,m+1})H(v_{m',m'-1})
\label{chipm}
\end{equation}
\begin{equation}
\chi^{\gamma}_{m,m-1}(u)\Phi^{\gamma}_{m',m'+1}(v)=\Theta(u_{m,m-1})\Theta(v_{m',m'+1})+H(u_{m,m-1})H(v_{m',m'+1})
\label{chimp}
\end{equation}
The arguments are
\begin{eqnarray}
u_{m,m+1}(u)-v_{m',m'+1}(v) = u-v+2(m+m')\eta+2t \nonumber \\
u_{m,m-1}(u)-v_{m',m'-1}(v) = u-v-2(n+n')\eta-2t \nonumber \\
u_{m,m+1}(u)-v_{m',m'-1}(v) = u-v+2(m-m'+1)\eta \nonumber \\
u_{m,m-1}(u)-v_{m',m'+1}(v) = u-v-2(m-m'-1)\eta \nonumber \\
u_{m,m+1}(u)+v_{m',m'+1}(v) = u+v+2(m-m')\eta\nonumber \\
u_{m,m-1}(u)+v_{m',m'-1}(v) = u+v+2(-n+n')\eta \nonumber \\
u_{m,m+1}(u)+v_{m',m'-1}(v) = u+v+2(m+n')\eta+2t \nonumber \\
u_{m,m-1}(u)+v_{m',m'+1}(v) = u+v-2(n+m')\eta-2t\nonumber \\
\end{eqnarray}
To rewrite (\ref{chipp})-(\ref{chimp}) we use 
\begin{equation}
\Theta(u)\Theta(v)+H(u)H(v) = cf_{+}(u+v)g_{+}(u-v)
\label{TpH}
\end{equation}
\begin{equation}
\Theta(u)\Theta(v)-H(u)H(v) = cf_{-}(u+v)g_{-}(u-v)
\label{TmH}
\end{equation}
\begin{equation}
f_{+}(u) = H((iK'+u)/2)H((iK'-u)/2) ~~~~g_{+}(u) = H_1((iK'+u)/2)H_1((iK'-u)/2)
\label{fgp}
\end{equation}
\begin{equation}
f_{-}(u) = H_1((iK'+u)/2)H_1((iK'-u)/2) ~~~~g_{-}(u) = H((iK'+u)/2)H((iK'-u)/2)
\label{fgm}
\end{equation}
We need especially the following properties of $g_{\pm}$
\begin{equation}
 g_{\pm}(-u) = g_{\pm}(u) \hspace{ 0.5in} g_{\pm}(u+4K) = g_{\pm}(u) 
\label{g}
\end{equation}
After insertion of (\ref{TpH})-(\ref{fgm}) into  (\ref{chipp})-(\ref{chimp}) we get
\begin{equation}
\chi^{\gamma}_{m,m+1}(u)\Phi^{\gamma}_{m',m'+1}(v)=cf_{-}(u+v+2(m-m')\eta)g_{-}(u-v+2(m+m')\eta +2t)
\label{g1}
\end{equation}
\begin{equation}
\chi^{\gamma}_{m,m-1}(u)\Phi^{\gamma}_{m',m'-1}(v)=cf_{-}(u+v+2(m'-m)\eta)g_{-}(u-v-2(n+n')\eta -2t)
\label{g2}
\end{equation}
\begin{equation}
\chi^{\gamma}_{m,m+1}(u)\Phi^{\gamma}_{m',m'-1}(v)= cf_{+}(u+v+2(m+n')\eta+2t)g_{+}(u-v+2(m-m'+1)\eta)
\label{g3}
\end{equation}
\begin{equation}
\chi^{\gamma}_{m,m-1}(u)\Phi^{\gamma}_{m',m'+1}(v)= cf_{+}(u+v-2(n+m')\eta-2t)\eta)g_{+}(u-v-2(m-m'-1)\eta)
\label{g4}
\end{equation}
It now remains to show that a $L^2\times L^2$ matrix $Y$ exists such that 
equ. (\ref{YS}) is satisfied for $\s_R$ and $\s_L$. As $\tau$ and $\tau'$ occurring in the definition of $\s_R$ and $\s_L$ are free parameters
we obtain from (\ref{YS})
\begin{equation}
\chi^{\gamma}_{m,n}(u)\Phi^{\gamma}_{m',n'}(v)=Y_{{m,m'};{k,k'}}\chi^{\gamma}(v)_{k,l}\Phi^{\gamma}(u)_{k',l'}Y^{-1}_{{l,l'};{n,n'}}
\label{YY}
\end{equation}
Taking tentatively $Y$ to be diagonal
\begin{equation}
Y_{{m,m'};{k,k'}} = y_{m,m'}\delta_{m,k}\delta_{m',k'}
\end{equation}
we get
\begin{equation}
\chi^{\gamma}_{m,n}(u)\Phi^{\gamma}_{m',n'}(v)=\frac{y_{m,m'}}{y_{n,n'}}\chi^{\gamma}_{m,n}(v)\Phi^{\gamma}_{m',n'}(u)
\end{equation}
and it follows from (\ref{g1})-(\ref{g4})
\begin{equation}
y_{m+1,m'+1}=y_{m,m'}\frac{g_{-}(u-v-2(m+m')\eta-2t)}{g_{-}(u-v+2(m+m')\eta +2t)}
\label{ypp}
\end{equation}
\begin{equation}
y_{m+1,m'-1}=y_{m,m'}\frac{g_{+}(u-v-2(m-m'+1)\eta)}{g_{+}(u-v+2(m-m'+1)\eta))}
\label{ypm}
\end{equation}
To prove that a matrix $Y$ can be found such that (\ref{YY}) is satisfied we  have to show that 
the set of equations  (\ref{ypp})-(\ref{ypm}) is free from contradictions on the torus of size $L\times L$
where 
\begin{equation}
y_{m+L,n+L} = y_{m,n}
\end{equation}
It follows from equ. (\ref{ypp}) that
\begin{eqnarray}
&&y_{m+L,n+L} = \nonumber \\
&&          \frac{g_-(u-v-2(m+n)\eta-4(L-1)\eta-2t)}{g_-(u-v+2(m+n)\eta+4(L-1)\eta+2t)}
            \frac{g_-(u-v-2(m+n)\eta-4(L-2)\eta-2t)}{g_-(u-v+2(m+n)\eta+4(L-2)\eta+2t)}\cdots \nonumber \\
&&          \frac{g_-(u-v-2(m+n)\eta-2t)}{g_-(u-v+2(m+n)\eta+2t)}y_{m,n}
\label{per}
\end{eqnarray}
The factor $g_-(u-v-2(m+n)\eta+4r_2\eta-2t)$ in the numerator cancels the factor
$g_-(u-v+2(m+n)\eta+4r_1\eta+2t)$ in the denominator if $t=0$ and
\begin{equation}
-2(m+n)\eta-4r_2\eta = 2(m+n)\eta+4r_1\eta +  4kK
\end{equation}
for arbitrary $k$ and if we set $k = 2m_1k_1$ for integer $k_1$
\begin{equation}
r_2 = k_1L -m-n-r_1
\end{equation}
It follows that for each factor in the numerator of equ. (\ref{per}) there is a factor in the
denominator against which it cancels.
Similarly we derive from equ. (\ref{ypm}) that
\begin{equation}
y_{m,n} = y_{m-L,n+L}
\end{equation}
We have shown that all $y_{m,n}$ can be determined from a single element (e.g. $y_{1,1}$) consistently if
$t=0$. This conclusion cannot be drawn for $t\neq 0$. A numerical test of (\ref{QLQR}) shows that it is not
satisfied for $t\neq 0$ and therefore no similarity transformation (\ref{YS}) exists for $t \neq 0$.
\\
\\
We summarize what has been found in this section :\\
\begin{quote}
\em
We have attained our goal to construct a Q-matrix which exists for $\eta=2mK/L$ for odd L:
\\
The $\Q_R$-matrix defined in equ. (\ref{TrSR}) with local matrices $\s_R$ defined in (\ref{SRkl})
is regular.\\
If the parameter $t$ is set to zero relation (\ref{QLQR}) is satisfied.\\
 Then $\Q(v)=\Q_R(v)\Q^{-1}_R(v_0)$
satisfies equ. (\ref{funeqn}) and commutes with the transfer matrix $T$.
\end{quote}
\section{Quasiperiodicity properties of $Q$.}
\noindent
It is easily seen that $Q_{72,R}(v)$ and $\Q_R(v,t)$ satisfy
\begin{equation}
\Q_{72,R}(v+2K) = S\Q_{72,R}(v)   \hspace{0.6 in}  \Q_R(v+2K,t) = S\Q_R(v,t)
\label{per1}
\end{equation}
It is of great importance to find the quasiperiodicity properties of the Q-matrices in the complex $v$-plane.
We do that in this section for $Q_{72,R}(v)$ and $\Q_R(v,t=0)$.
It is well known that the quasiperiod of $Q_{73}$ is $i\Ks'$. See \cite{baxb} for details.
We shall present plausibility arguments for the statement that $Q_{72,R}$ as well as $\Q$ are singular matrices
if their quasiperiod is $i\Ks'$.

\subsection{Quasiperiodicity properties of $Q_{72}$.}
\noindent
We get from equs. (\ref{SR}), (\ref{qperH}), (\ref{qperT}) and from
 \begin{equation}
\eta = m_1K/L
\end{equation}
the relations
\begin{eqnarray} 
\begin{array}{lccl} 
  {\Ss_R(\pm,\beta)_{k,k+1}}(v+i\Ks') & = & & f(v)\exp(+i\pi k\eta/K)\Ss_R(\mp,\beta)(v)_{k,k+1}\\
  {\Ss_R(\pm,\beta)_{k+1,k}}(v+i\Ks') & = & & f(v)\exp(-i\pi k\eta/K)\Ss_R(\mp,\beta)(v)_{k+1,k}\\
  {\Ss_R(\pm,\beta)_{1,1}}(v+i\Ks')   & = & & f(v)\Ss_R(\mp,\beta)(v)_{1,1}\\
  {\Ss_R(\pm,\beta)_{L,L}}(v+i\Ks')   & = &  (-1)^{m_1}&f(v)\Ss_R(\mp,\beta)(v)_{L,L}\\
\end{array}
\end{eqnarray}
where 
\begin{equation}
f(v) = q^{-1/4}\exp(-\frac{i\pi v}{2K})
\end{equation}
The similarity transformation
\begin{equation}
  {\bs(\alpha,\beta)_{i,l} = A_{i,j} \s(\alpha,\beta)_{j,k} A^{-1}_{k,l}}
\end{equation}
with
\begin{equation}
  {A_{k,l} = \delta_{k,l}\exp(\frac{i\pi}{2K}(k-1)k\eta) a_1}
\end{equation}
leads to
\begin{eqnarray} 
\begin{array}{lccl} 
  {\tSs_R(\pm,\beta)_{k,k+1}}(v+i\Ks') & = & & f(v)\Ss_R(\mp,\beta)(v)_{k,k+1}\\
  {\tSs_R(\pm,\beta)_{k+1,k}}(v+i\Ks') & = & & f(v)\Ss_R(\mp,\beta)(v)_{k+1,k}\\
  {\tSs_R(\pm,\beta)_{1,1}}(v+i\Ks')   & = & & f(v)\Ss_R(\mp,\beta)(v)_{1,1}\\
  {\tSs_R(\pm,\beta)_{L,L}}(v+i\Ks')   & = &  (-1)^{m_1}&f(v)\Ss_R(\mp,\beta)(v)_{L,L}\\
\end{array}
\end{eqnarray}
If $m_1$ is even it follows that
\begin{equation}
{\tSs_R(\alpha,\beta)_{k,l}}(v+i\Ks') = f(v)R(\alpha,\gamma)\Ss_R(\gamma,\beta)(v)_{k,l}
\end{equation}
where $R$ is defined in equ. (\ref{SRop}).
and 
\begin{equation}
Q_{R,72}(v+i\Ks') = f(v)^{N} R Q_{R,72}(v)
\label{QR72qp}
\end{equation}
However it is well known \cite{Newdev} that for even $m_1$ $Q_{R,72}(v)$ is singular and
consequently relation (\ref{QR72qp}) cannot be upgraded from  $Q_{R,72}$ to $Q_{72}$.
It is shown in \cite{Newdev} that instead of (\ref{QR72qp}) the following relation holds
\begin{equation}
Q_{R,72}(v+2i\Ks') =  q^{-N}\exp(-iN\pi v/K) Q_{R,72}(v)
\label{QR72qp2Kp}
\end{equation}
which is correct for all $\eta=m_1K/L$.
Provided $m_1$ is odd it follows
\begin{equation}
Q_{72}(v+2i\Ks') =  q^{-N}\exp(-iN\pi v/K) Q_{72}(v)
\label{Q72qp2Kp}
\end{equation}
\subsection{Quasiperiodicity properties of $\Q$.}
We obtain from equ. (\ref{SRkl}).
\begin{equation}
\s_R(\pm,\beta)(v+i\Ks')_{k,k+1} = f(v)(-i)\exp(+i\pi k\eta/K)\s_R(\mp,\beta)(v)_{k,k+1}
\end{equation}
\begin{equation}
\s_R(\pm,\beta)(v+i\Ks')_{k+1,k} = f(v)(+i)\exp(-i\pi k\eta/K)\s_R(\mp,\beta)(v)_{k+1,k}
\end{equation}
Perform the similarity transformation
\begin{equation}
  {\bs(\alpha,\beta)_{i,l} = A_{i,j} \s(\alpha,\beta)_{j,k} A^{-1}_{k,l}}
\end{equation}
with
\begin{equation}
  {A_{k,l} = \delta_{k,l}(-i)^{k-1}\exp(\frac{i\pi}{2K}(k-1)k\eta) a_1}
\end{equation}
Then for $k<L$
\begin{equation}
\bs_R(\pm,\beta)(v+i\Ks')_{k,k+1} = f(v)\s_R(\mp,\beta)(v)_{k,k+1}
\end{equation}
\begin{equation}
\bs_R(\pm,\beta)(v+i\Ks')_{k+1,k} = f(v)\s_R(\mp,\beta)(v)_{k+1,k}
\end{equation}
and for $k=L$
\begin{equation}
\bs_R(\pm,\beta)(v+i\Ks')_{L,1} = f(v)\exp\left[\frac{+i\pi}{2}((2m_1-1)L+2m_1)\right]\s_R(\mp,\beta)(v)_{k,k+1}
\end{equation}
\begin{equation}
\bs_R(\pm,\beta)(v+i\Ks')_{1,L} = f(v)\exp\left[\frac{-i\pi}{2}((2m_1-1)L+2m_1)\right]\s_R(\mp,\beta)(v)_{k+1,k}
\end{equation}
We find that if
\begin{equation}
\exp\left[\frac{-i\pi}{2}((2m_1-1)L+2m_1)\right] = 1
\label{condR}
\end{equation}
$\Q_R$ satisfies the relation
\begin{equation}
\Q_R(v+i\Ks') = q^{-N/4}\exp(-\frac{i\pi Nv}{2K}) R \Q_R(v)
\label{QRqper}
\end{equation}
which is the same as (\ref{QR72qp}) for $Q_{R,72}$.
This happens only \\
I.)for even $m_1$ if $L=4 \times$ integer\\
II.)for odd  $m_1$ if $L=2\times$ odd integer.\\
These are exactly those cases in which $\Q_R$ is singular.
Like equ. (\ref{QR72qp}) equ. (\ref{QRqper}) does not give the corresponding relation for
the Q-matrix $\Q$. \\

In the following paragraph $\Qs$ denotes either $Q_{72}$ or $\Q$.

\begin{quote}
\em
\noindent
We note that if the relation 
\begin{equation}
\Qs(v+i\Ks') = q^{-N/4}\exp(-\frac{i\pi Nv}{2K}) R \Qs_(v)
\label{Qqper}
\end{equation}
were correct then it would follow that 
\begin{equation}
q(v+i\Ks') |q> =  q^{-N/4}\exp(-\frac{i\pi Nv}{2K})q(v) R|q>
\end{equation}
where $|q>$ denotes an arbitrary eigenvector of $\Qs(v)$ and $q(v)$ its eigenvalue.\\
In other words : all eigenvectors of $\Q_(v)$ would be eigenvectors of $R$.
It is however well known \cite{Newdev} that the eigenvectors of $Q_{72}(v)$ which are degenerate
eigenvectors of the transfer matrix T are generally not eigenvectors of R.
\end{quote}
Equs. (\ref{QR72qp}) and (\ref{QRqper}) allow a coherent explanation of the fact that $Q_R$ is singular for
one set of $\eta$ values and regular for another.
Under the assumption that if $\Qs$ exists there are eigenstates of $\Qs$
which are not eigenstates of R $Q_R$ cannot be regular if in case of $Q_{72}$ $m_1$ is even
or in case of $\Q$ (\ref{condR}) is satisfied. This explains also naturally the observation
that for fixed L and sufficiently small N $Q_R^{-1}$ exists always as then all states are singlets
and  (\ref{QR72qp}) and (\ref{QRqper}) do not lead to contradictions when upgraded from $\Qs_R$ to $\Qs$.
\\
Using the method used in this section it can easily be shown that always
\begin{equation}
\Q_R(v+2iK') = q^{-N} \exp(-iN\pi v/K)\Q_R(v)
\label{perR2}
\end{equation}
and consequently if $\Q_R^{-1}$ exists
\begin{equation}
\Q(v+2iK') = q^{-N} \exp(-iN\pi v/K)\Q(v)
\label{perQ2}
\end{equation}
\section{The properties of $\Q$ for $t=0$.}
\noindent
It follows from (\ref{perQ2}) that as shown for $Q_{72}$ in  \cite{Newdev} $\Q(v)$ may be written as
\begin{eqnarray}
\Q(v)=\hat{{\cal K}}(q;v_k){\rm exp}(i(n_B-\nu)\pi v/2K)\prod_{j=1}^{n_B}
h(v-v^B_j)\nonumber\\
\times\prod_{j=1}^{n_L}H(v-iw_j)H(v-iw_j-2\eta)\cdots H(v-iw_j-2(L-1)\eta)
\label{form2}
\end{eqnarray}
\begin{equation}
2n_B+Ln_L=N.
\label{norootsa}
\end{equation}
$n_B$ is the number of Bethe roots $v^{B}_k$ and $n_L$ the number of exact $Q$-strings of length $L$.
The sum rules (2.16)-(2.21) of ref.\cite{Newdev} are also true for $\Q(v)$.\\
From (\ref{per1}) and 
\begin{equation}
\Q_L(v)\Q_R(u)=\Q_L(u)\Q_R(v)
\label{newQLQR}
\end{equation}
follows that 
\begin{equation}
[S,\Q(v)] = 0
\label{SQ}
\end{equation}
Finally we find numerically that the functional relation (\ref{con}) which was originally
conjectured in ref. \cite {Newdev} is also satisfied for $\Q(v)$.
\section{The matrix $\Q(v,t)$.}
\noindent
We have shown that $\Q_R(v)$ satisfies the $T\Q_R$ relation and found that it is
not singular. But the proof of relation (\ref{QLQR}) failed for parameter $t \neq 0$.  
One finds numerically that (\ref{QLQR}) is in fact violated for systems large enough to allow degenerate eigenvalues
of the transfer matrix.
Therefore the question arises whether $\Q_R(v,t)$ is useful at all. Surprisingly we find numerically that
for $\eta=2mK/L$ and odd $L$ despite
\begin{equation}
\Q^{-1}_L(v_0,t)\Q_L(v,t) \neq \Q_R(v,t)\Q^{-1}_R(v_0,t)
\label{QneqQ}
\end{equation}
both matrices
\begin{equation}
\Q^{(L)}(v,t) = \Q^{-1}_L(v_0,t)\Q_L(v,t)~~~~~{\rm and}~~~~~ \Q^{(R)}(v,t) = \Q_R(v,t)\Q^{-1}_R(v_0,t)
\label{QRL}
\end{equation}
commute with the transfer matrix $T$.
Furthermore we find that in the cases studied $\Q^{(L)}(v,t)$ and $\Q^{(R)}(v,t)$ have the same eigenvalues.
This means that there exists a matrix $A$ with $\Q^{(L)}(v,t)=A\Q^{(R)}(v,t)A^{-1}$ and consequently
instead of (\ref{QLQR})
\begin{equation}
\Q_L(v)A\Q_R(u)=\Q_L(u)A\Q_R(v)
\label{QLAQR}
\end{equation}
should hold.
A consequence of (\ref{QneqQ}) is that
\begin{equation}
[\Q^{(R)}(v_1,t),\Q^{(R)}(v_2,t)] \neq 0
\label{Qv1Qv2}
\end{equation}
as one needs (\ref{QLQR}) to prove that $Q$-matrices with different arguments commute (see 9.48.41 in 
ref. \cite{baxb}). 
We find that like $Q_{72}$ the matrix $\hat{Q}$ does not commute with $R$
\begin{equation}
[R,\hat{Q}(v,t)] \neq 0
\label{RQ}
\end{equation}
but unlike $Q_{72}$  as a consequence of (\ref{QneqQ}) does also not commute with $S$ for $t \neq 0$:
\begin{equation}
[S,\hat{Q}(v,t)] \neq 0
\label{SQt}
\end{equation}
This is possible because the degenerate subspaces of $T$ have elements with both eigenvalues
$\nu'=0,1$ of $S$ if \mbox{$\eta = 2m_1K/L$} and $L$ is odd.\\
These properties of $\Q^{(L)}(v,t)$ and $\Q^{(R)}(v,t)$ imply that they act as non abelian symmetry
operators in all degenerate subspaces of the set of eigenvectors of $T$.
We finally mention that whereas the six-vertex limit of $Q_{R,72}$ does not exist it exists for  $\hat{Q}_R$.
The limit of $\Q_R(v,t=iK'/2)$ for elliptic nome $q \rightarrow 0$ is well defined.
Using 
\begin{equation}
\lim_{q \rightarrow 0} H(u\pm iK'/2) = \exp(\mp i(u-\pi/2))   ~~~~~~
\lim_{q \rightarrow 0} \Theta(u\pm iK'/2) = 1
\end{equation}
one gets a regular limiting $\hat{Q}_R$-matrix.
It has been checked numerically that the resulting $\hat{Q}$-matrix commutes with $T_{6v}$. \\
\\
\\
\vspace{.4in}
{ \large Acknowledgement}\\
I am pleased to thank Prof. Barry M. McCoy for helpful comments and suggestions.

\vspace{.2in}

% .................................................................................................
\app{}
The transfer matrix of the eight vertex model is 
\begin{equation}
T(v)|_{ \mu,\nu}={\rm Tr} W_8(\mu_1,\nu_1)W_8(\mu_2,\nu_2)
\cdots W_8(\mu_N,\nu_N)
\label{t8}
\end{equation}
where in the conventions of  (6.2) of ref. \cite{bax0}
\begin{eqnarray}
W_8(1,1)|_{1,1}=W_8(-1,-1)|_{-1,-1}&=a=&\rho
\Theta(2\eta)\Theta(v-\eta)
H(v+\eta)\nonumber\\
W_8(-1,-1)|_{1,1}=W_8(1,1)|_{-1,-1}&=b=&\rho\Theta(2\eta)
H(v-\eta)\Theta(v+\eta)\nonumber\\
W_8(-1,1)|_{1,-1}=W_8(1,-1)|_{-1,1}&=c=&\rho H(2\eta)
\Theta(v-\eta)\Theta(v+\eta)\nonumber\\
W_8(1,-1)|_{1,-1}=W_8(-1,1)|_{-1,1}&=d=&\rho H(2\eta)
H(v-\eta)H(v+\eta).\nonumber\\
\label{bw}
\end{eqnarray}
Relations used in the text. See e.g. \cite{WW}
\begin{equation}
 \sn(u-v)=\frac{\sn(u) \cn(v) \dn(v) - \sn(v) \cn(u) \dn(u)}{1 - k^2 \sn^2(u) \sn^2(v)}
\label{snumv}
\end{equation}
\begin{equation}
H(v+iK') = iq^{-1/4}\exp(-\frac{i\pi v}{2K})\Theta(v)
\label{qperH}
\end{equation}
\begin{equation}
\Theta(v+iK') = iq^{-1/4}\exp(-\frac{i\pi v}{2K})H(v)
\label{qperT}
\end{equation}

\end{document}